# Predicting Cognitive Assessment Scores in Older Adults with Cognitive Impairment Using Wearable Sensors

**Assma Habadi**
Department of Electrical and Computer Engineering
University of Illinois Chicago
Chicago, IL, USA
Email: ahabad2@uic.edu

**Miloš Žefran, Ph.D.**
Department of Electrical and Computer Engineering
University of Illinois Chicago
Chicago, IL, USA

**Lijuan Yin, Ph.D.**
Institute for Health Research and Policy
University of Illinois Chicago
Chicago, IL, USA

**Woojin Song, Ph.D.**
Department of Neurology and Rehabilitation
University of Illinois Chicago College of Medicine
Chicago, IL, USA

**Maria Caceres, LPN**
Institute for Health Research and Policy
University of Illinois Chicago
Chicago, IL, USA

**Elise Hu**
School of Public Health, College of Medicine
University of Illinois Chicago
Chicago, IL, USA

**Naoko Muramatsu, Ph.D.**
School of Public Health and Institute for Health Research and Policy
University of Illinois Chicago
Chicago, IL, USA

**Abstract**

**Background and Objectives:** This paper focuses on using AI to assess the cognitive function of older adults with mild cognitive impairment or mild dementia using physiological data provided by a wearable device. Cognitive screening tools are disruptive, time-consuming, and only capture brief snapshots of activity. Wearable sensors offer an attractive alternative by continuously monitoring physiological signals. This study investigated whether physiological data can accurately predict scores on established cognitive tests.

**Research Design and Methods:** We recorded physiological signals from 23 older adults completing three NIH Toolbox Cognitive Battery tests, which assess working memory, processing speed, and attention. The Empatica EmbracePlus, a wearable device, measured blood volume pulse, skin conductance, temperature, and movement. Statistical features were extracted using wavelet-based and segmentation methods. We then applied supervised learning and validated predictions via cross-validation, hold-out testing, and bootstrapping.

**Results:** Our models showed strong performance with Spearman's $\rho$ of 0.73–0.82 and mean absolute errors of 0.14–0.16, significantly outperforming a naive mean predictor. Sensor roles varied: heart-related signals combined with movement and temperature best predicted working memory, movement paired with skin conductance was most informative for processing speed, and heart in tandem with skin conductance worked best for attention.

**Discussion and Implications:** These findings suggest that wearable sensors paired with AI tools such as supervised learning and feature engineering can noninvasively track specific cognitive functions in older adults, enabling continuous monitoring. Our study demonstrates how AI can be leveraged when the data sample is small. This approach may support remote assessments and facilitate clinical interventions.

# 1 Background and Objectives

Artificial Intelligence (AI) holds great promise for improving the lives of older adults. It has been used to monitor older adults for falls or adverse events and alert caregivers, provide reminders about medications and medical appointments, allow older adults to connect with friends or family and avoid isolation, provide decision support for caregivers and healthcare providers, and create social robots that provide companionship and other services. The focus of this paper is on the use of AI for diagnosis and personalization of healthcare. In particular, our focus is on developing novel ways for cognitive assessment of older adults with mild cognitive impairment (MCI) or mild dementia.

Standardized neuropsychological tests inform dementia diagnosis by providing quantitative measures of an individual's cognitive function within and across multiple domains. Comprehensive neuropsychological evaluation typically takes 6-8 hours. Reducing the time required to administer tests while providing quantitative measures of key cognitive functions has been a major area of research efforts, especially within clinical research settings. NIH Toolbox Cognitive Battery (NIH-TB-CB) is one of these instruments and was recently developed to assess multiple domains of cognition in an efficient manner utilizing digital technology (Weintraub et al., 2014). While individual NIH-TB-CB tests only last minutes, they must be administered in person by trained staff using an iPad. The primary goal of our study was to understand if physiological markers measured by a wearable device as adults aged 50 or over with a diagnosis of MCI or mild dementia were taking an NIH-TB-CB test can predict their test score. Using physiological signals measured by a wearable device for cognitive assessment could allow assessment during everyday activities. In turn, this could enable early detection of cognitive impairment and would open the door for a host of novel clinical interventions. In addition, despite the large literature on neuropsychological tests, little is known about how older adults experience neuropsychology assessment tests. To gain some insight into this question we investigated which physiological signals were correlated with the scores on different tests, and thus with different types of cognitive activities.

MCI is a condition that is estimated to affect around 15–20% of older adults (60 or older); it is the bridge between normal aging and dementia (Petersen, 2016). Cognitive screening tools like Mini-Mental State Examination (MMSE) and Montreal Cognitive Assessment (MoCA) enable clinicians and care teams to implement therapeutic measures that help delay disease progression while maintaining patient independence (Livingston et al., 2020). They are administered in person and provide a snapshot of cognitive functioning at the time of assessment (Nasreddine et al., 2005; Tsoi et al., 2015). However, older adults show performance variability over time due to fatigue and changes in their circadian rhythms (Hood et al., 2017).

Wearable technology systems that operate through continuous monitoring produce extended time-series data about cognitive health indicators (Kourtis et al., 2019). A variety of research studies demonstrate how machine learning (ML) algorithms detect cognitive impairment markers by analyzing motor patterns, circadian rhythms, and gait and physiological signals (Rykov et al., 2024; Liu et al., 2022; Li et al., 2023). The combination of observational data with ML algorithms proved successful in distinguishing MCI patients from control subjects (Seifallahi et al., 2024; Xu et al., 2024; Liu et al., 2022). For instance, Xu et al., (2024) achieved perfect MCI subject identification using supervised ML classifiers that combined physical activity, heart rate, and sleep data. The supervised ML classifiers demonstrated 85% accuracy when identifying MCI patients through gait data collected during clinical gait assessments (Seifallahi et al., 2024).

Additional neuro-cognitive tests may explain hidden signs of cognitive decline beyond conventional diagnostic categories of MCI and normal cognition. Sakal et al. (2024) retrieved data from National Health and Nutrition Examination Survey databases, which studied wearable device data from older adults who took three cognitive tests measuring processing speed, attention, and working memory. Using a decision tree ensemble model, for one of the tests the study achieved median AUC results that exceeded 0.82 when predicting poor versus high cognitive performance, highlighting robust discriminative ability. Similarly, a clinical study of older adults with amnestic MCI employed a detailed neuropsychological test battery for evaluation. The supervised ML regression analysis of wearable device signals produced predictions for executive function, processing speed, immediate and delayed memory, and global cognition test scores, which achieved up to Spearman's $\rho = 0.69$ with true results (Rykov et al., 2024).

The assessment of cognitive function through wearable sensor data and ML appears feasible based on recent positive findings. Yet, researchers face essential obstacles to measuring the specific relationship between physiological signals and cognitive function. The majority of research conducted focused on basic categorization of participants into MCI versus normal cognition groups or poor versus high performers groups (Xu et al., 2024; Sakal et al., 2024). In order to better personalize care, finer-grained measures of cognitive performance are necessary. The analysis of physiological signals linked to these measures could provide additional insights and improve the ability of clinicians to monitor the progression of the cognitive decline. Finally, the reported results would be further strengthened if they could be replicated with the recently developed cognitive assessment tools that simplify the assessment, such as NIH-TB-CB (Weintraub et al., 2014).

The reliability of the outcomes of the existing studies remains a critical issue since most single-arm clinical trials enroll limited participants ($N = 12–61$) because recruiting patients from this specific population group is difficult, which presents a major challenge for ML methods (Seifallahi et al., 2024; Liu et al., 2022; Xu et al., 2024; Rykov et al., 2024; Li et al., 2023). Although some validation approaches have been proposed in the current literature, there remains a need for more rigorous accuracy and robustness testing to confirm these findings in a replicable manner.

This research examined the capability of noninvasive wearable technology to forecast NIH-TB-CB scores among older adults who received MCI or mild dementia diagnoses. The analysis focused on three specific cognitive measures: working memory, processing speed, and attention. The wearable device collected physiological signals which were subsequently processed by specialized AI pipelines employing supervised learning regression models that are tuned for each specific cognitive test. Our study is characterized by small data rather than the more common big-data applications of AI. The specialized AI pipelines are particularly well suited for small data settings. An ablation study examined each pipeline to determine which physiological signals provide the most significant contribution to cognitive test score predictions. In turn, this allowed us to gain insight into how older adults experience these cognitive tests. Multiple accuracy and robustness tests were carried out to confirm the validity of the obtained results. This research fills a knowledge void by applying AI and ML to forecast and verify NIH-TB-CB test scores in MCI or mild dementia patients. It is important to point out that in contrast to existing studies that use long sequences (hours or days) of data, our work allows near real-time (minutes) assessment of cognitive function.

## 2  Research Design and Methods

### 2.1  Participants and Procedures

The research was performed within a randomized controlled trial which tested the feasibility of incorporating a gentle physical activity program adapted for older patients aged 50 and above diagnosed with MCI or mild dementia at a Memory & Aging Clinic (MeC).  The study included patients who could maintain a seated position independently for 15 minutes, had decision-making abilities for consent, were able to speak English , and were physically inactive with less than 150 minutes of planned physical exercise per week. The study excludes participants who have major physical or mental disabilities, who participate in other physical activity research, or who experience uncorrectable hearing or vision problems.

A total of 161 potential participants were identified between March 2023 and December 2024 via clinical records or through neuropsychological evaluations during their MeC visit. Only 28 of these participants were enrolled in this study, highlighting the difficulties of recruitment for this population. Although the sample is relatively small, such data is especially valuable given the population.

All research activities for the study were conducted at the MeC and its designated research zones within the building. All participants were scheduled to complete research visits twice, before and after the intervention. Each visit consisted of a survey interview, physical performance tests, and cognitive assessment using NIH-TB-CB. As an optional component, the participants who consented wore the Empatica EmbracePlus medical-grade device (U.S. Food & Drug Administration, 2022) as a wrist-worn sensor to measure physiological signals during both visits. The University of Illinois Chicago Institutional Review Board authorized the research protocol (including Empatica EmbracePlus testing). The clinic staff explained study procedures and risks and benefits to participants after obtaining their consent in a private examination area. The study participants provided written consent either at their clinic appointment or at their initial research appointment when additional time was required to make decisions.

### 2.2 Data Collection

The Empatica EmbracePlus recorded four important signals: (1) Blood Volume Pulse (BVP) measured through optical sensors which detect microvascular blood pulsation changes in absorption levels at 64 Hz sampling rate. The measured data are stored as nanowatts. (2) Electrodermal Activity (EDA) sampled at 4 Hz and recorded in units of microsiemens ($\mu$S). (3) Temperature recorded using a thermistor operating at 1 Hz to measure skin temperature, which produces results in degrees Celsius (°C). (4) Acceleration data recorded at 64 Hz from the three axes $x$, $y$ and $z$. The device's Inertial Measurement Unit provides both physical and digital range information for each axis, recording data as integers via analog-to-digital conversion. Empatica uses Amazon Web Services to store deidentified raw sensor files after implementation of server-side encryption based on HIPAA and institutional review board (IRB) requirements.

This study utilized five NIH-TB-CB tests: (1) The Flanker Inhibitory Control and Attention Test (Attention Test) measures attention and inhibitory control by requiring participants to indicate the direction of a target while ignoring distracting stimuli. (2) The List Sorting Working Memory Test (Working Memory Test) measures working memory through tasks that present verbal and pictorial item lists that participants need to remember and repeat according to specific criteria (e.g., from smallest to largest). (3) The Dimensional Change Card Sort Test (Executive Function Test) measures cognitive flexibility and attention by presenting the participant with two pictures and then asking them to match a third picture along one of two factors (e.g., shape and color), with the matching factor determined by the computer and presented as a word before the third picture. (4) The Pattern Comparison Processing Speed Test (Processing Speed Test) assesses the processing speed through a task that requires participants to verify the similarity of paired simple images within a limited time periods. (5) The Picture Sequence Memory Test (Episodic Memory Test) measures episodic memory by flexibly presenting a 'story' via words and pictures and then asking the participant to put the pictures in the same order as they were just shown.  The five NIH-TB-CB tests have an approximate completion time of 3, 7, 4, 3, and 7 minutes. It should be noted that the test completion duration varies from one participant to another, depending on their individual abilities. During each test session, the Empatica EmbracePlus collects physiological data that will be used to predict the participants' NIH-TB-CB test scores. Although five NIH-TB-CB tests were administered, due to time constraints and the limited scope of this study, only the data from the Working Memory Test, Processing Speed Test, Attention Test, and Episodic Memory Test were analyzed. The NIH-TB-CB scores were automatically provided by the app. The uncorrected standard scores were used in this study.

By capturing each participant's physiological signals in real time and linking them to scores on the NIH-TB-CB tests, we generate a rich dataset for exploring both cognitive performance and potential stress responses. Changes in these signals may indicate how older adults are experiencing the demands of each test.

### 2.3  Preprocessing and Feature Engineering

Processing methods designed to improve sensor data quality and maintain uniformity across modalities were applied to all signals recorded by the Empatica EmbracePlus device. Artifact detection and removal algorithms identified and excluded anomalous values arising from sensor misplacement or movement artifacts (i.e., noise segments in which data are not physiologically valid) (Clifford et al., 2011). Following artifact cleaning, tailored filtering techniques were used: for BVP, a bandpass filter isolated the pulsatile component of the cardiovascular signal; for EDA, a low-pass filter smoothed high-frequency noise and drift (Oppenheim, 1999; Proakis, 2001; Smith et al., 1997). Accelerometry data underwent a bandpass filter to capture meaningful motion while minimizing low-frequency drift and high-frequency interference. A moving average filter was used to smooth skin temperature data minimizing transient fluctuations (Clifford et al., 2006). Then, z-scoring (standardizing each sensor to have mean 0 and standard deviation 1) was used on each sensor to ensure all features would be on comparable scales (James et al., 2013), enabling multi-modal integration for predictive modeling.

Two feature extraction methods, wavelet-based and segment-based, were applied to the physiological data. The wavelet-based approach decomposed each physiological signal into multiple frequency bands (using wavelet transforms) to capture both persistent low-frequency trends and short-lived, transient events. This multi-resolution analysis is especially suitable for signals like BVP and EDA, which can exhibit nonstationary behavior across multiple frequency domains (Stiles et al., 2004). The effectiveness of wavelet-based methods in capturing nonstationary signal features has been recognized in the literature (Addison, 2005; Mallat, 1999).

The segment-based method divided the continuous signals into a fixed number of quasi-stationary segments, from which statistical descriptors were extracted. Dividing the signals into segments can reveal localized variations that might be missed in entire-signal analyses (Naqvi et al., 2020). Both the wavelet-based and segment-based approaches extracted equivalent statistical measurements (energy, mean, standard deviation, minimum, maximum, skewness, and kurtosis) but did so at different granularities (frequency bands vs. time segments). To determine which approach better captures task-specific physiological patterns, the features of each cognitive test's physiological data have been extracted using wavelet-based extraction, segment-based extraction, or a combination of both. This allowed an independent assessment of each technique's ability to detect meaningful patterns correlated with cognitive performance.

The use of wavelet-based and segment-based feature extraction methods allowed the evaluation of temporal and frequency-domain characteristics of the physiological signals. A thorough examination of different wavelet levels, wavelet functions, and number of segmentations, creates a solid system to determine optimal features for cognitive performance prediction.

### 2.4 Predictive Modeling

Building on the feature set derived from the physiological signals, we employ predictive modeling approaches that are well suited to small-sample contexts with high-dimensional, potentially noisy data. The chosen modeling approach includes two main categories: (1) regularized linear models (Lasso, Ridge, and Elastic Net), and (2) a Random Forest-based two-stage system. These choices reflect our data's characteristics: correlated physiological features combined with a limited number of observations.

The characteristics of our physiological signal features (many predictors, few samples) make regularized linear models especially relevant. Lasso regression (L1-penalized) can drive coefficients for less informative features to zero, effectively performing variable selection in high-dimensional contexts (Tibshirani, 1996; Finch & Finch, 2016). Ridge regression (L2-penalized) shrinks all coefficients but does not force them to zero, which is beneficial when predictors are correlated (Chen et al., 2020). Elastic Net unites both L1 and L2 penalties in one framework, balancing variable selection with coefficient shrinkage. We use cross-validation

(repeatedly splitting the data into training and validation sets) to choose regularization parameters, thereby preventing overfitting with limited data (Zou & Hastie, 2005; Zhang et al., 2021). Collectively, these methods handle correlated features effectively while retaining interpretability (Zhang et al., 2021).

In addition to regularized models, we explored a hybrid two-stage strategy to address the heterogeneous and nonlinear characteristics inherent in physiological responses. In the first stage, a Random Forest Classifier was trained to categorize samples as "low-score" or "high-score," using the median score in the training fold as a threshold. A Random Forest is an ensemble-learning method that combines multiple decision trees to capture more complex patterns in the data (Breiman, 2001; Strobl et al., 2009). The classifier outputs a probability ($p_{low}$) for each sample that reflects the model's confidence in assigning it to the low-score group. This leverages the nonlinear relationships often seen in physiological signals (He et al., 2024). In the second stage, we trained two distinct Random Forest Regressors to model the subgroups separately: one for the low-score group ($RF_{low}$) and another for the high-score group ($RF_{high}$). Each regressor specialized in capturing subgroup-specific relationships.

Based on the probability ($p_{low}$), samples with ($p_{low} < \tau_L$) or ($p_{low} > \tau_U$) are routed to either $RF_{high}$ or $RF_{low}$. For borderline cases ($p_{low} \in [\tau_L, \tau_U]$), the final prediction is a weighted blend of each regressor's output. (Detailed equations are provided in Section S1 of the Supplementary Material). By partitioning the data into more homogeneous subsets, this two-stage approach reduced heterogeneity and enhanced predictive accuracy (Han et al., 2021; Wozniak & Zahabi, 2024).

Generally, both the regularized linear models and the hybrid two-stage approach offer robust solutions to the challenges posed by high-dimensional, noisy physiological signals and a limited number of samples. These models minimize overfitting risks while allowing us to isolate features or subgroups that best predict cognitive performance in older adults with mild dementia or MCI.

### 2.5 Model Selection and Evaluation

We performed a grid search over the feature extraction strategies (wavelet vs. segmentation), wavelet parameters (e.g., decomposition levels and wavelet types), segment counts, and model types (Lasso, Ridge, Elastic Net, and the Two-stage Random Forest). Targeted parameter tuning further refined the chosen combination, producing the best predictions. A full sensitivity analysis of these parameters for each cognitive test appears in Section S2 of the Supplementary Material.

The main evaluation employed is the repeated 5-fold cross-validation. In 5-fold cross-validation, the dataset is partitioned into 5 subsets (folds); for each iteration, one-fold is used for testing, and the remaining 4 folds are used for training, and this process is repeated so every fold serves as a test fold once (Kohavi et al., 1995). The process was repeated 5 times with different random splits to provide more robust performance estimates. We reported Mean Absolute Error (MAE) to gauge average prediction error, Spearman's $\rho$ to assess monotonic relationships, and Pearson correlation to measure linear relationships. MAE was computed for each fold and then averaged. Out-of-fold predictions for each sample were aggregated to calculate Spearman's $\rho$ and Pearson correlations. An early pruning mechanism halted a given hyperparameter configuration if partial results exceeded a predefined MAE margin above the best observed performance.

Multiple robustness checks supplemented the 5-fold cross-validation. A hold-out test set, meaning a portion of data reserved solely for final testing, provided additional evidence of generalizability. Leave-One-Out Cross-Validation (LOOCV) was also conducted, wherein each sample is left out exactly once as a test example, using all other data for training. We further employed bootstrapping (resampling the data with replacement many times) to estimate the variability of MAE and correlations, and we performed a permutation test (randomly shuffling the outcome values many times to assess whether the observed correlations might arise by chance) to examine the statistical significance of the correlation metrics (Good, 2013). All hypothesis tests used $p < .05$ as the significance threshold. We also compared results against a naive predictor that always outputs the mean training score.

A sensor ablation study was also conducted to evaluate the relative contribution of each sensor modality. Features associated with a specific sensor were systematically removed, and predictive performance was compared with and without those sensor inputs. This procedure allowed us to identify redundant versus critical sensor signals for cognitive performance prediction.

All preprocessing, feature extraction, predictive modeling, and evaluations were conducted in Python 3.9 using NumPy, Pandas, SciPy, PyWavelets, and scikit-learn (VanderPlas, 2016; Virtanen et al., 2020).

### 3 Results

#### 3.1 Participant Characteristics

The majority of participants ($\approx 61\%$) were female and aged between 50 and 90 years. Among these participants, 23 were introduced to the Empatica EmbracePlus device and all of them consented to wearing it during their visits. Two of these participants did not attend their follow-up appointments, and one was introduced to the wearable device only during the follow-up visit. At the time of analysis, two participants had not yet completed their follow-up visits, and a technical issue prevented the successful administration of the NIH-TB-CB follow-up assessment for one participant. Thus, for this analysis, we included data from 22 baseline sessions and 18 follow-up sessions.

During every clinic visit, participants wore the Empatica EmbracePlus device nonstop. The NIH-TB-CB tests were administered without any notable data loss from sensor detachment or technical errors, suggesting that continuous physiological data collection in this population is highly feasible. The scores from the NIH-TB-CB tests were collected as planned, except for two participants who were given no score for their Working Memory Test because they failed the practice questions. The differences in test lengths and performance results among these evaluations are compiled in Table 1.

### 3.2 NIH-TB-CB Scores Prediction

In this section, the findings of our predictive modeling on the three NIH-TB-CB tests are presented, detailing which physiological signals best predict each test score. Since certain physiological signals can be linked to arousal or stress responses (Cacioppo et al., 2016), their predictive strength may reflect the degree of bodily stress that the older adults experience during different cognitive tasks.

We investigated two feature extraction techniques (wavelet-based versus segment-based) and multiple model types (Lasso, Ridge, Elastic Net, and Two-stage Random Forest), as detailed in the Research Design and Methods section. Only the top-performing configurations for each test are presented here, despite the fact that other combinations were examined. We used grid search to find a good starting set of features and model settings, and then we tuned specific parameters to make those configurations even better.

#### 3.2.1 List Sorting Working Memory Test

For the Working Memory Test, we implemented a segment-based feature extraction approach using 5 segments and combined the resulting features with Ridge regression, which yielded the best performance using BVP, temperature, and $x$-axis from the accelerometer. A brief exploration of the model settings, including different values of the Ridge parameter, is provided in Table S1 in section S2.1 in the Supplementary Material.

Table 2 row "Working Memory" compares the Spearman's $\rho$ of this model's configuration against a naive (mean-based) prediction method (which always predicts the average training score) under four validation methods (5×5 k-fold, Hold-Out, LOOCV, Bootstrapping). The naive predictor had a negatively correlated value of $-0.293$, denoting a weak predictive ability. By contrast, this best model achieves substantially higher positive correlations under all validation strategies, with a maximum value of Spearman's $\rho = 0.822$ with 5×5 k-fold validation. The model upholds its solid performance levels (0.519–0.764) across different validation split methods.

Similarly, Table 3 shows that the naive predictor's MAE of 0.233 decreases to as low as 0.136 under LOOCV and reaches 0.143 with the 5×5 k-fold cross-validation. This shows a sizable gain in prediction accuracy, demonstrating consistent out-performance compared to naive prediction.

Having established the model's superior performance across various validation methods, we next examine the individual sensor contributions to understand which modalities caused these improvements. Figure 1A reveals that the BVP sensor combined with the accelerometer produces a high positive correlation ($\rho \approx 0.51$), whereas EDA and temperature demonstrate a weak negative relationship ($\rho \approx -0.15$). These values suggest that the combined use of BVP and accelerometry signals shows effective synergy in predicting these Working Memory Test scores, while temperature and EDA signals provide minimal additional value.

The boxplots of sensor ablation in Figure 2A confirm these results as it shows that including the $x$-axis accelerometer and BVP produces significantly stronger Spearman's $\rho$ than omitting them. The performance enhancement from $y$-axis, $z$-axis, EDA, and temperature measurements remain small or non-existent compared to other variables. Overall, these results reinforce that $x$-axis accelerometry and BVP are the main contributors to predictive performance for the Working Memory Test.

### 3.2.2 Pattern Comparison Processing Speed Test

For the Processing Speed Test, we found that Lasso regression, combined with the wavelet-based features, delivered the most effective results when used with EDA and accelerometer signals. A more detailed overview of the tested Lasso settings for this model can be found in Table S2 in section S2.2 of the Supplementary Material.

In Table 2 (row "Processing Speed"), the Spearman's $\rho$ of this model is compared to that of a naive (mean-based) predictor across multiple validation methods. The naive predictor, with a score of $-0.325$ exhibited a weak performance, whereas the proposed methods consistently produced higher positive correlations that reach a maximum value of 0.881 in the hold-out split evaluation. This Lasso-based approach proved superior by producing higher correlations regardless of validation methods. Table 3 confirms these findings by showing that the naive predictor's MAE (0.212) drops to a range of 0.159–0.194 when employing the optimized wavelet-Lasso configuration.

Looking at the sensors' modalities contributions, the accelerometer demonstrates notably stronger positive correlation than other sensors. Specifically combining accelerometer with EDA in which it achieved ($\rho \approx 0.73$) and with BVP resulting in ($\rho \approx 0.70$) according to the sensor synergy heatmap Figure 1B. The accelerometer's diagonal correlation of ($\rho \approx 0.71$) is also noteworthy. By contrast, BVP, EDA, and temperature maintain mild to moderate negative correlations among themselves, including negative correlations between BVP and EDA and BVP and temperature range from mild to moderate at $-0.16$ and $-0.28$ respectively, indicating minimal shared predictive value.

Looking deeper into the accelerometer components, Figure 2B shows that the inclusion of $y$ and $z$ accelerometer axes lead to higher achieved Spearman's $\rho$ when compared to omitting these axes. The same figure shows that the impact of temperature and BVP is often negligible or slightly adverse, while EDA shows overlapping distributions with modest improvements at best. According to these observations, the Processing Speed Test benefits consistently from accelerometer data while the other sensor inputs show less predictable effects.

### 3.2.3 Flanker Inhibitory Control and Attention Test

The initial grid search revealed that a two-stage pipeline with wavelet-based feature extraction was the best setup for forecasting the Attention Test scores, specifically when using the BVP, temperature, $x$-axis, and EDA signals. Table S3 in section S2.3 in the Supplementary Material provides the complete list of hyperparameter settings (e.g., blending weight, number of trees) evaluated for this two-stage pipeline.

In this two-stage framework, probability thresholds were set at $\tau_L = 0.3$ and $\tau_U = 0.4$. Instances with $p_{low} < 0.3$ were routed to the high-score regressor, those above 0.4 went to the low-score regressor, and cases falling between these bounds used a weighted blend of both regressors.

Row "Attention" in Table 2 compares the strongest Spearman's $\rho$ from this model to a naive baseline under multiple validation scenarios. The proposed approach achieves higher predictive accuracy than the baseline with a hold-out set correlation of 0.810 and 0.734 under 5×5 k-fold cross-validation. At the same time, the naive predictor shows a weaker negative correlation of -0.558. This two-stage pipeline also reduces MAE substantially as Table 3 indicates that the naive predictor 0.317 error level drops to between 0.104 and 0.231, which demonstrates better prediction accuracy.

Turning to the sensor's synergy in Figure 1C, BVP and EDA demonstrate the highest correlation of ($\rho \approx 0.51$) while EDA shows low correlation of ($\rho \approx 0.04$) on the diagonal. Temperature and accelerometer each demonstrate moderate correlations with other sensors of around 0.25–0.39 and 0.30–0.37, respectively, indicating that no other pairing matches BVP–EDA's high positive relationship.

Figure 2C further clarifies each sensor's individual impact. Removing the $x$, $y$, or $z$ accelerometer axes produce correlations similar to or slightly above including them, suggesting accelerometer data may not be pivotal here. In contrast, temperature and EDA yield modest gains, while BVP provides the largest boost (its median correlation rises from about 0.2 to 0.4 when included). The inclusion of temperature and EDA sensor also yield some improvement. Overall, these findings indicate that temperature, EDA, and especially BVP enhance Attention Test score predictions, whereas the accelerometer axes contribute little under the current setup.

### 3.2.4 Picture Sequence Memory Test

The grid search for the Episodic Memory Test didn't yield any configuration that achieved good results. All of the predicted results failed to attain a Spearman's $\rho$ correlation above the threshold typically considered meaningful (e.g., 0.3), and its MAE did not show a substantial improvement (defined here as at least a 20% reduction) over that of a naive predictor. Therefore, no further results are reported for this test.

## 4  Discussion and Implications

The primary objective of this experimental study was to determine whether AI tools such as supervised learning and feature engineering could use physiological data provided by a wearable device to reliably predict NIH-TB-CB cognitive test scores in older adults diagnosed with MCI or mild dementia. Overall, the results indicate that physiological signals obtained from the Empatica EmbracePlus device capture meaningful information about an individual's cognitive performance, as the predictive models consistently outperformed naive baselines across these three targeted domains

We find that cognitive abilities, such as Working Memory, Processing Speed, and Attention, are represented in different physiological and motor cues captured on the wrist. More specifically, this expands on previous research, which has primarily studied the broad classification of MCI versus control status (Seifallahi et al., 2024; Xu et al., 2024; Sakal et al., 2024). Our research goes beyond simple binary diagnosis and suggests that modern AI techniques can use physiological measurements provided by wearable devices to identify more nuanced, domain-relevant changes in cognitive abilities, which is a critical step toward early detection and tracking of cognitive decline. The sensor ablation study further analyzed the contributions of different physiological markers. It showed that BVP and accelerometer features are highly predictive of the Working Memory test score, while accelerometer and EDA features were the best in predicting a Processing Speed score. The combination of BVP and EDA signals yielded the best performance for the Attention Test. These findings underscore that distinct physiological processes may reflect different aspects of cognition, and no single sensor modality universally dominates every task. The results of our analysis on the Episodic Memory Test did not yield significant predictive results. A plausible reason for that is that, as seen in Table 1, the score distribution of this test was much narrower than in the other tests (the standard deviation for this test was ±7.15; for the others, it ranged between 13.64 and 17.87). This narrower score range might be an indication that the NIH-TB-CB Episodic Memory Test is less capable of capturing subtle differences in cognitive performance in this population. However, this hypothesis requires further investigation.

Several plausible psychophysiological links may explain why some signals are more prominent. BVP measures heart rate-derived indicators that are known to align with changes in an individual's autonomic nervous system that occur in response to cognitive load and stress situations (Chen et al., 2020; Zhang et al., 2021). Another observation is that signals of heart rate and heart rate variability become more prominent while performing sustained attention and rapid decision-making tasks, which parallels our finding in BVP and EDA signal synergy in the prediction of Attention Test. On the other hand, accelerometer and EDA data are frequently related to fine-grained motor output, which is important for tasks requiring quick or sustained responses, like the Processing Speed Test (Xu et al., 2024; He et al., 2024).

The comprehensive feature extraction strategy is another methodological highlight. On the one hand, wavelet-based methods proved to be better suited for analyzing nonstationary physiological signals in the Attention and Processing Speed tasks, while segment analysis was effective in the Working Memory Test. Results from previous work suggest that wavelet transforms excel at capturing transient or short-lived patterns in autonomic responses (Addison, 2005; Stiles et al., 2004; Mallat, 1999). In contrast, segment-based approaches (assuming quasi-stationarity) may better capture ongoing physiological states during longer tasks with sustained cognitive demands (Naqvi et al., 2020). These results validate that our signal processing pipelines must be customized for each cognitive domain rather than working within a one-size-fits-all approach.

From a clinical perspective, the study's demonstration of continuous, noninvasive monitoring of older adults with cognitive impairment is promising. Unlike traditional screening tests, wearable technology can capture repeated, within-person physiological data using continuous monitoring, providing repeated, within-person snapshots of cognitive status across the day. This is possible because modern AI techniques can be leveraged to classify various activities throughout the day and identify periods corresponding to a particular cognitive activity. Cognitive assessment through wearable technology would not only ease the burden of frequent in-person testing (Kourtis et al., 2019; Rykov et al., 2024), it could be used to more readily assess interventions aimed at improving MCI or mild dementia, especially those that target nonclinical settings. This could dramatically expand the space for new interventions and simplify their development.

This study's strengths include demonstrating that older adults with MCI or mild dementia can wear a medical-grade device that collects multiple physiological signals comfortably without sensor detachment or data loss. Furthermore, we showed the feasibility and acceptability of such wearable devices for this population as all the 23 participants introduced to the Empatica EmbracePlus device consented to wear it. IN contrast to existing studies, the data was collected during a relatively short period when the person took a specific test thus capturing the physiological response directly triggered by the test. The study is further unique because it focuses on the NIH-TB-CB, which guarantees that the physiological responses coincide precisely with validated cognitive domains. We also ran additional validation and robustness evaluation tests on our results, including k-fold cross-validation, LOOCV and bootstrapping. To further validate our findings, a permutation test was also performed. This suite of tests is essential for small datasets to prevent overfitting with robust statistical verification (Han et al., 2021; Wozniak & Zahabi, 2024).

Despite that, there are limitations that should be addressed in future work. For example, we limited our focus to only four of these five tests because of time constraints and the scope of the study itself. Future research will incorporate the analysis of the Executive Function Test data to further expand the scope of cognitive activities that can be assessed using our approach. Another major limitation of our study is the data size. As outlined in section 2.1, we identified 161 potential participants but only enrolled 28. Of these 28, 23 were introduced to the Empatica EmbracePlus device and all of them consented to wear it. This relatively low enrollment rate showcases the challenges with recruitment for this population and highlights that each participant's data is extremely valuable. However, this limits the complexity of the models that can be reliably trained. Although effective in reducing overfitting, regularization techniques (Lasso, Ridge, Elastic Net) and ensemble methods can also limit our findings' generalizability. Additionally, the short time frame for data collection during each cognitive test limits the ability to observe potentially emerging patterns of physiological data that might be observed over longer or more naturalistic time frames. Repeated assessments in home and community settings would better define how circadian rhythms, mood states, and everyday stress influence physiological cognitive contingencies.

Despite these limitations, the study still presents a robust framework and proof of concept for wearable device-based prediction of domain-specific cognitive performance. Our findings provide a strong rationale for future multi-sensor integration and task-optimized feature extraction by showing that each sensor modality contributes uniquely to Working Memory, Attention, and Processing Speed. Further extension of these efforts will require a deeper examination of which specific features within each sensor domain most reliably map onto cognitive states. Understanding these sensor features may shed light on the biological mechanism of mapping sensor signals onto cognitive abilities to support more informed clinical interventions. Future studies should incorporate a more diverse population and longer naturalistic monitoring periods to validate these sensor-based insights and leverage wearable devices as key tools for more proactive and personalized care.

**Funding**

This work was partially supported by the National Institute on Aging of the National Institutes of Health [Grant number P30AG022849] and National Science Foundation [Grant number IIS1705058]. We acknowledge the support of the UIC Center for Clinical and Translational Science for REDCap, funded by the National Center for Advancing Translational Sciences, National Institutes of Health, through Grant UL1TR002003. The content is solely the responsibility of the authors and does not necessarily represent the official views of the National Institutes of Health.

**Conflict of Interest**

None declared.

**Acknowledgments**

no acknowledgments to report

**Table and Figure Captions**

Table 1. Completion Time and Performance Scores for Cognitive Tests.

Table 2. Robustness Evaluation of Best-Performing Model Configurations for each NIH-TB-CB Test (Spearman's $\rho$).

Table 3. Robustness Evaluation of Best-Performing Model Configurations for each NIH-TB-CB Test (MAE).

Figure 1. Pairwise sensor synergy heatmaps (Spearman's $\rho$) for three cognitive tasks—(A) List Sorting Working Memory, (B) Pattern Comparison Processing Speed, and (C) Flanker Inhibitory Control and Attention. Each heatmap encompasses four sensor modalities (BVP, EDA, Temperature, Accelerometer). Diagonal cells represent performance using a single modality, while off-diagonal cells represent performance when combining the two modalities at their intersection.

Figure 2. Paired boxplots showing the distribution of Spearman's $\rho$ for each sensor across ablation trials in three cognitive tests: (A) List Sorting Working Memory, (B) Pattern Comparison Processing Speed, and (C) Flanker Inhibitory Control and Attention. For each sensor on the y-axis, the two boxplots compare all trial runs in which that sensor was included in the model (blue)—whether by itself or together with other sensors—versus excluded from the model (green).

**Tables/Figures**

Table 1: Completion Time and Performance Scores for Cognitive Tests

| Test | Completion Time (Minutes) (±SD) | Score (±SD) |
|---|---|---|
| List Sorting Working Memory Test | 7.90 ± 1.72 | 81.89 ± 15.48 |
| Pattern Comparison Processing Speed Test | 3.43 ± 0.57 | 78.30 ± 17.87 |
| Flanker Inhibitory Control and Attention Test | 4.50 ± 0.75 | 81.22 ± 13.64 |
| Picture Sequence Memory Test | 9.45 ± 3.04 | 83.75 ± 7.15 |
| Dimensional Change Card Sort Test | 5.71 ± 0.67 | 88.22 ± 15.36 |

Table 2: Robustness Evaluation of Best-Performing Model Configurations for each NIH-TB-CB Test (Spearman's $\rho$)

| Cognitive Domain | Feature Extraction Method | ML Model | Sensor Modalities | Naive Predictor | 5×5-fold | Hold-Out | LOOCV | Bootstrapping |
|---|---|---|---|---|---|---|---|---|
| Working Memory | Segment-based | Ridge | BVP, Accelerometer ($x$-axis), Temperature | -0.293 | 0.822 | 0.638 | 0.764 | 0.519 |
| Processing Speed | Wavelet-based | Lasso | Accelerometer ($x$-axis, $y$-axis, $z$-axis), EDA | -0.325 | 0.731 | 0.881 | 0.573 | 0.495 |
| Attention | Wavelet-based | Two-Stage Random Forest | BVP, EDA, Temperature, Accelerometer ($x$-axis) | -0.558 | 0.734 | 0.810 | 0.610 | 0.509 |

Table 3: Robustness Evaluation of Best-Performing Model Configurations for each NIH-TB-CB Test (MAE)

| Cognitive Domain | Feature Extraction Method | ML Model | Sensor Modalities | Naive Predictor | 5×5-fold | Hold-Out | LOOCV | Bootstrapping |
|---|---|---|---|---|---|---|---|---|
| Working Memory | Segment-based | Ridge | BVP, Accelerometer ($x$-axis), Temperature | 0.233 | 0.143 | 0.151 | 0.136 | 0.195 |
| Processing Speed | Wavelet-based | Lasso | Accelerometer ($x$-axis, $y$-axis, $z$-axis), EDA | 0.212 | 0.159 | 0.164 | 0.168 | 0.194 |
| Attention | Wavelet-based | Two-Stage Random Forest | BVP, EDA, Temperature, Accelerometer ($x$-axis) | 0.317 | 0.158 | 0.104 | 0.193 | 0.231 |

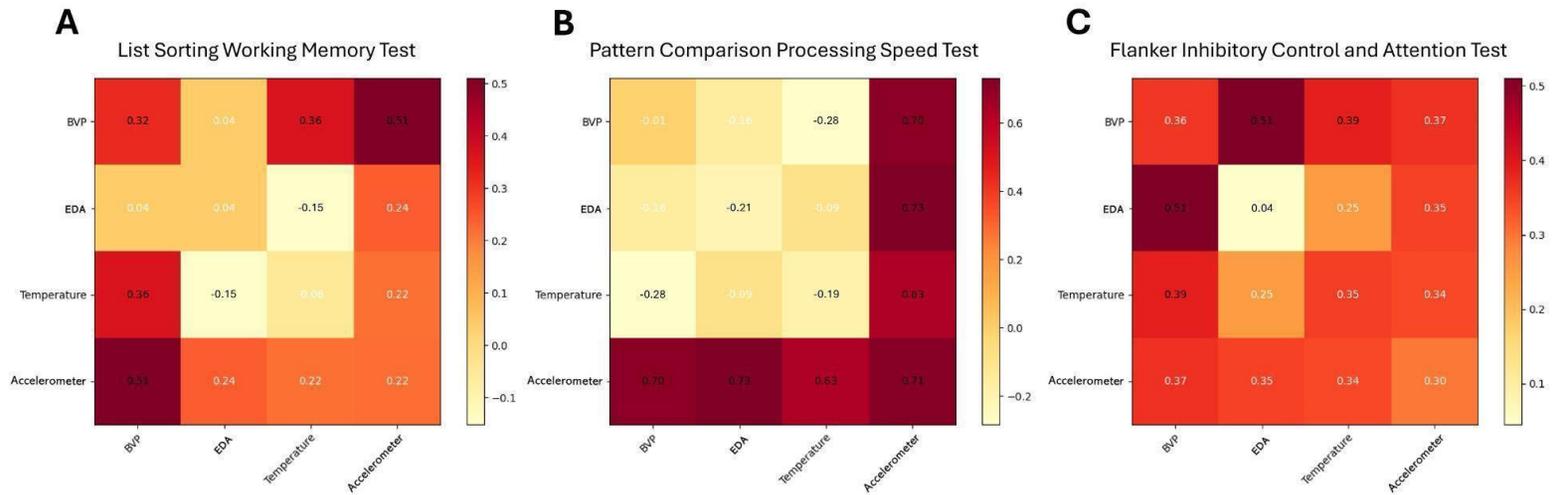

Figure 1: Pairwise sensor synergy heatmaps (Spearman's $\rho$) for three cognitive tasks—(A) List Sorting Working Memory, (B) Pattern Comparison Processing Speed, and (C) Flanker Inhibitory Control and Attention. Each heatmap encompasses four sensor modalities (BVP, EDA, Temperature, Accelerometer). Diagonal cells represent performance using a single modality, while off-diagonal cells represent performance when combining the two modalities at their intersection.

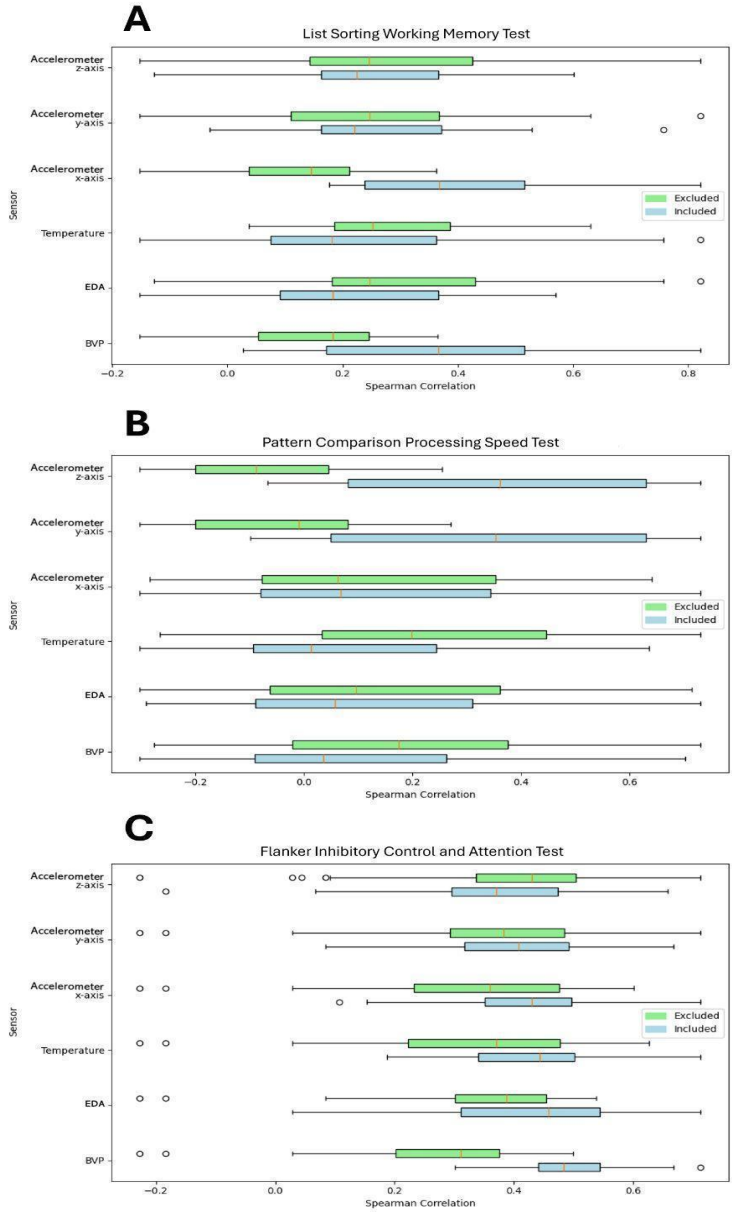

Figure 2: Paired boxplots showing the distribution of Spearman's $\rho$ for each sensor across ablation trials in three cognitive tests: (A) List Sorting Working Memory, (B) Pattern Comparison Processing Speed, and (C) Flanker Inhibitory Control and Attention. For each sensor on the y-axis, the two boxplots compare all trial runs in which that sensor was included in the model (blue)—whether by itself or together with other sensors—versus excluded from the model (green).

**Supplementary Material**

**S1. Equations for the Two-Stage Random Forest Blending**

We adopt a hybrid two-stage strategy, starting with a Random Forest Classifier that outputs a probability $p_{low}$ indicating the likelihood a sample belongs to the low-score group. Next, two Random Forest Regressors ($RF_{low}$ and $RF_{high}$) specialize in their respective subgroups.

- If $p_{low} < \tau_L$, use $RF_{high}$ exclusively.
- If $p_{low} > \tau_U$, use $RF_{low}$ exclusively.

When $p_{low}$ lies between $\tau_L$ and $\tau_U$, the final prediction $\hat{y}$ is a weighted blend:

$$\alpha = \frac{1}{2}\left(\omega + \frac{p_{low} - \tau_L}{\tau_U - \tau_L}\right)$$

$$\hat{y} = \alpha \cdot \hat{y}_{low} + (1 - \alpha)\,\hat{y}_{high}$$

where $\omega$ is a blending hyperparameter tuned during cross-validation, and $\tau_L, \tau_U$ are fixed thresholds.

## S2. Detailed Hyperparameter Tuning Results

### S2.1. List Sorting Working Memory Test

A segment-based feature extraction with 5 segments and Ridge regression yielded the best performance. Table 4 shows how different values of the regularization parameter $\alpha$ affected performance. We chose $\alpha$ values spanning roughly half, near, and double the optimum found by cross-validation to confirm robustness around that optimal point. These results reflect a sensitivity analysis across $\alpha$. Sensor modalities included: BVP, temperature, $x$-axis accelerometer.

Table S1: Sensitivity analysis for the Working Memory Test using Ridge regression under varying $\alpha$.

| $\alpha$ | MAE ($\pm$SD) | Spearman's $\rho$ | Pearson |
|---|---|---|---|
| 0.88 | 0.140 $\pm$ 0.035 | 0.822 | 0.812 |
| 0.45 | 0.144 $\pm$ 0.034 | 0.819 | 0.807 |
| 0.60 | 0.142 $\pm$ 0.035 | 0.819 | 0.809 |
| 1.10 | 0.139 $\pm$ 0.036 | 0.818 | 0.813 |
| 1.80 | 0.137 $\pm$ 0.038 | 0.816 | 0.813 |

### S2.2. Pattern Comparison Processing Speed Test

A wavelet-based approach using Daubechies 6 at decomposition level 3 combined with Lasso regression performed best. Table 5 shows how varying $\alpha$ influences accuracy. Values were chosen to explore a wide range (including roughly half and double the optimum) so we could verify that performance remains consistent if $\alpha$ slightly deviates from the best estimate. Sensor modalities included: EDA, $x$-axis, $y$-axis, $z$-axis accelerometer.

Table S2: Sensitivity analysis for the Processing Speed Test using Lasso regression under varying $\alpha$.

| $\alpha$ | MAE ($\pm$SD) | Spearman's $\rho$ | Pearson |
|---|---|---|---|
| 1.58 | 0.159 $\pm$ 0.032 | 0.731 | 0.697 |
| 0.80 | 0.174 $\pm$ 0.449 | 0.618 | 0.648 |
| 1.10 | 0.166 $\pm$ 0.038 | 0.669 | 0.684 |
| 2.00 | 0.159 $\pm$ 0.029 | 0.686 | 0.676 |
| 3.20 | 0.171 $\pm$ 0.029 | 0.584 | 0.608 |

## S2.3. Flanker Inhibitory Control and Attention Test

For the Attention Test, a two-stage Random Forest approach with wavelet-based extraction (db4, level 4) was optimal. Table 6 presents the main hyperparameters tested: $\omega$ (the blending weight), the number of trees in the Random Forest Classifier ($N_{cls}$), and in the Random Forest Regressors ($N_{reg}$). We systematically varied each parameter over a practical range to find a near-optimal balance of MAE and correlation. Sensor modalities included: BVP, temperature, $x$-axis accelerometer, EDA.

Table S3: Sensitivity analysis for the Attention Test using a two-stage Random Forest system, varying $\omega$, $N_{cls}$, and $N_{reg}$.

| $\omega$ | $N_{cls}$ | $N_{reg}$ | MAE ($\pm$SD) | Spearman's $\rho$ | Pearson |
|---|---|---|---|---|---|
| 0.50 | 155 | 40 | 0.158 $\pm$ 0.047 | 0.734 | 0.763 |
| 0.30 | 155 | 40 | 0.160 $\pm$ 0.047 | 0.727 | 0.740 |
| 0.70 | 155 | 40 | 0.163 $\pm$ 0.041 | 0.724 | 0.766 |
| 0.50 | 105 | 40 | 0.177 $\pm$ 0.045 | 0.680 | 0.713 |
| 0.50 | 205 | 40 | 0.171 $\pm$ 0.043 | 0.632 | 0.712 |
| 0.50 | 155 | 20 | 0.160 $\pm$ 0.045 | 0.716 | 0.752 |
| 0.50 | 155 | 60 | 0.166 $\pm$ 0.046 | 0.704 | 0.750 |